\documentclass[journal]{IEEEtran}

\usepackage{amsmath,amssymb,amsfonts,bm}

\usepackage{graphicx}
\usepackage{tikz}
\usepackage{circuitikz}
\usepackage{dblfloatfix}
\usepackage{float}
\usetikzlibrary{shapes.geometric, arrows.meta, positioning, calc, fit, backgrounds, shadows}

\usepackage{booktabs}
\usepackage{tabularx}
\usepackage{threeparttable}
\usepackage{siunitx}

\usepackage{cite}
\usepackage[hidelinks]{hyperref}

\usepackage{textcomp}
\usepackage{xcolor}

\newcommand{\NP}{N_P}
\newcommand{\NL}{N_L}
\newcommand{\Cdim}{C}
\newcommand{\vect}[1]{\bm{#1}}
\newcommand{\mat}[1]{\mathbf{#1}}

\begin{document}

\title{Resource-Efficient CSI Prediction: A Gated Fusion and Factorized Projection Approach}

\author{Mohammad~Hussain,
        Maedeh~Adibag, Dilara~Gurer, Gokhan~Kalem, Kerim~Serin,
        and~Sinem~Coleri,~\IEEEmembership{Fellow,~IEEE}%
\thanks{M. Hussain, M. Adibag, and S. Coleri are with the Department of Electrical and Electronics Engineering, Ko\c{c} University, Istanbul, Turkey (email: \{mohussain25, mbag24, scoleri\}@ku.edu.tr). D. Gurer and G. Kalem are with Turkcell (email: \{dilara.gurer, gokhan.kalem\}@turkcell.com.tr). K. Serin is with Opticoms (email: kerim.serin@opticoms.de). This work is supported by the Scientific and Technological Research Council of Turkey (TUBITAK) 1711 Project AI-PG5 \#3247019.}%
}

\maketitle

\begin{abstract}

Accurate Channel State Information (CSI) prediction is essential for dynamic multiple-input multiple-output (MIMO) systems but remains computationally demanding. This letter proposes a resource-efficient predictor that combines a gated recurrent unit (GRU) encoder with Luong attention, a bottleneck gated fusion module, and a Dimension-wise Separable Linear Head (DSLH). The gated fusion module integrates local recurrent features with global attention context, while the DSLH reduces the cost of the output mapping. Evaluated on 3GPP TR~38.901-compliant channels, the proposed model achieves an average NMSE of $-13.84$~dB with 26\% fewer parameters and approximately $2.3\times$ higher inference throughput than a dimension-matched LinFormer baseline. The proposed model is best suited to LOS and mixed-condition scenarios, offering a practical accuracy--efficiency trade-off for short-horizon CSI prediction at moderate sequence lengths.
\end{abstract}

\begin{IEEEkeywords}
Channel prediction, gated recurrent unit (GRU), MIMO, resource-efficient deep learning, throughput efficiency.
\end{IEEEkeywords}

\vspace{-0.5em}
\section{Introduction}

\IEEEPARstart{T}{he} performance of MIMO systems relies heavily on accurate CSI. In dynamic wireless environments, the channel varies rapidly, causing CSI to ``age" between acquisition and use, leading to severe performance degradation. To mitigate this, deep learning (DL) predictors have been widely adopted to forecast future CSI.

Early DL approaches treated CSI matrices as visual images and employed Convolutional Neural Networks (CNNs) to extract spatial patterns. Vilas Boas \emph{et al.}~\cite{TypeII_NR} leveraged this idea for CSI compression, while~\cite{evoCSINet} extended it to 3D-CNNs to capture temporal correlations. Although CNNs model local spatial features effectively, they struggle to capture long-range temporal dependencies, which are important for accurate channel prediction in dynamic environments.

Recurrent neural networks (RNNs) such as long short-term memory (LSTMs) and gated recurrent units (GRUs) better exploit temporal dynamics through hidden-state memory~\cite{NeuralProphet2022, SmartCSI_VTC2023}, with recent work further improving GRU-based CSI prediction via EVT-based adaptive loss functions~\cite{MehrniaCL2025}. 
However, standard RNNs process sequences strictly sequentially, which limits parallelization and can amplify error propagation over long horizons.

Transformer-based predictors address RNN limitations including limited parallelization, poor scalability to long sequences, and error accumulation through self-attention, enabling parallel processing over the full history. Jiang \emph{et al.}~\cite{JiangJSAC2022} showed that global attention mitigates mobility-related degradation, while physics-inspired variants such as ODE-Former~\cite{ODEFormer2025} model fading in continuous time. However, transformers incur quadratic attention cost $O(N^2)$, and even efficient variants such as LinFormer~\cite{LinFormer2025} retain large embeddings, leading to high parameter and computational overhead that is unsuitable for real-time systems~\cite{CompNN2024}.

Prior recurrent and attention-based architectures for CSI prediction already exist, including dual-attention LSTMs for industrial IoT \cite{Hakimi2025}, adaptive bidirectional GRUs for underwater MIMO \cite{Hu2024}, attention-enhanced GRUs for deep-space channels \cite{deep}, and dynamic graph neural networks for massive MIMO \cite{Kumar2025}. Rather than proposing GRU--attention itself as a new idea, this paper focuses on a resource-efficient CSI prediction architecture that combines a GRU encoder, Luong-style attention, bottleneck gated fusion, and a DSLH output head, and evaluates this combination under standardized 3GPP conditions. The resulting design targets short-horizon prediction with a favorable balance between accuracy, model size, and inference efficiency. The proposed architecture uses bottleneck gated fusion to adaptively combine encoder states with global attention context before the DSLH head. Compared to fixed linear fusion, the learned gate is more flexible, while DSLH provides a compact alternative to a dense output head.
The specific contributions of this letter are as follows:
\begin{itemize}
\item \emph{Bottleneck gated fusion:} A lightweight module that combines each time-step feature with a shared attention context using a learned gate. Results show a 0.66 dB improvement over simple linear fusion.
\item \emph{DSLH with recurrent encoders:} We show that DSLH works well with recurrent models when attention and gated fusion are used to improve the sequence features. This allows efficient multi-step prediction without needing a large dense output layer.
\item \emph{3GPP accuracy--efficiency evaluation:} The proposed
model achieves best mean NMSE in LOS and mixed-condition
scenarios with 26\% fewer parameters and $2.3\times$ higher
throughput than LinFormer, evaluated across five 3GPP
TR~38.901 scenarios.
\end{itemize}
The remainder of this letter is organized as follows: Section~\ref{sec:system_model} details the system model. Section~\ref{sec:method} describes the proposed architecture. Section~\ref{sec:results} presents the experimental evaluation, including ablation studies, complexity analysis, and accuracy assessments, in various 3GPP scenarios. Section~\ref{sec:conclusion} concludes the paper.

\vspace{-0.5em}
\section{System Model}\label{sec:system_model}

\subsection{MIMO Channel and Notation}
We consider a downlink narrowband MIMO channel observed at discrete time index $t$,
\begin{equation}
\mat{H}_t \in \mathbb{C}^{R \times T},
\end{equation}
where $T$ and $R$ denote the number of transmit and receive antennas, respectively.
We focus on a single representative subcarrier.

The channel is modeled using a path-resolved representation. Let $\mat{H}_{t,p} \in \mathbb{C}^{R \times T}$ denote the
contribution of path $p \in \{1,\ldots,P\}$ at time $t$, where $P$ is the total number of multipath components. The effective
(path-collapsed) channel is defined as
\begin{equation}
\mat{H}^{(\mathrm{eff})}_t \triangleq \sum_{p=1}^{P} \mat{H}_{t,p}.
\end{equation}
\subsection{Input and Output Representation}
At each time $t$, the effective channel matrix
$\mat{H}^{(\mathrm{eff})}_t$ is converted to a real-valued representation by
vectorizing and stacking its real and imaginary components:
\begin{equation}
\vect{h}^{(\mathrm{r})}_t \triangleq
\begin{bmatrix}
\Re\{\operatorname{vec}(\mat{H}^{(\mathrm{eff})}_t)\} \\
\Im\{\operatorname{vec}(\mat{H}^{(\mathrm{eff})}_t)\}
\end{bmatrix}
\in \mathbb{R}^{2RT}.
\end{equation}
To incorporate mobility information, the scalar UE speed $v_t$ is appended to form the per-time-step input feature vector:
\begin{equation}
\vect{x}_t =
\big[
(\vect{h}^{(\mathrm{r})}_t)^{\top},\;
v_t
\big]^{\top}
\in \mathbb{R}^{D}, \qquad D \triangleq 2RT + 1.
\end{equation}

Given the past $\NP$ observations, the model input is the windowed tensor
\begin{equation}
\mat{X}_t \in \mathbb{R}^{\NP \times D}.
\end{equation}

We predict future channel \emph{increments} over a horizon of $\NL$ steps.
Specifically, increments are defined in the complex channel domain as
\begin{equation}
\Delta \mat{H}^{(\mathrm{eff})}_{t+k}
\triangleq
\mat{H}^{(\mathrm{eff})}_{t+k}
-
\mat{H}^{(\mathrm{eff})}_{t+k-1},
\qquad k=1,\dots,\NL.
\label{eq:increment}
\end{equation}
Each increment is represented in real-valued vectorized form for learning.
Accordingly, the predictor outputs the multi-step increment tensor
\begin{equation}
\mat{Y}_t \in \mathbb{R}^{\NL \times \Cdim},
\qquad \Cdim \triangleq 2RT.
\end{equation}
Increment prediction reduces the dynamic range by removing slowly varying
components and stabilizes learning.

\section{Proposed GRU--Attention--DSLH Predictor}\label{sec:method}
The proposed architecture combines a GRU encoder, Luong-style attention, 
and bottleneck gated fusion, followed by a DSLH that maps the 
refined sequence to the $\NL$-step prediction horizon.
\begin{figure}[htbp]
 \centering
 \hspace*{2cm}
\begin{tikzpicture}[
  scale=0.9,
  transform shape,
  node distance=1.0cm,
  arrow/.style={thick, -Stealth, color=black},
  baseblock/.style={
rectangle,
rounded corners=3pt,
minimum width=3.0cm,
minimum height=0.8cm,
text centered,
draw=black,
thick,
font=\small\sffamily
  },
  process/.style={baseblock, fill=blue!10},
  attention/.style={baseblock, fill=green!10},
  projection/.style={baseblock, fill=orange!15},
  container/.style={draw=black!50, thick, rounded corners=5pt, dashed, inner sep=0.4cm}
 ]

 \node (inputMath) [font=\footnotesize, align=center] {
  \textbf{Input} $\mathbf{X}_t \in \mathbb{R}^{N_P \times D}$
 };

 \node (gru) [process, above=0.9cm of inputMath] {GRU Encoder};

 \node (attn) [attention, above=0.9cm of gru] {Luong Attention + Gated Fusion};

 \node (timeProj) [projection, above=1.1cm of attn] {Time Projection $\mathbf{W}_{time}$};
 \node (trans) [process, minimum width=2.5cm, fill=gray!10,
 above=0.8cm of timeProj] {Transpose};
 \node (chanProj) [projection, above=0.8cm of trans] {Channel Projection $\mathbf{W}_{ch}$};

 \node (outputMath) [above=1.0cm of chanProj, font=\footnotesize, align=center] {
  \textbf{Target} $\mathbf{Y}_t \in \mathbb{R}^{N_L \times C}$
 };

 \begin{pgfonlayer}{background}
  \node[container, fit=(timeProj) (trans) (chanProj)] (dslhbox) {} ;
 \end{pgfonlayer}

 \node[font=\bfseries\small, text=gray, anchor=north west]
  at ([xshift=0.1cm,yshift=-0.05cm] dslhbox.north west) {DSLH};

 \draw[arrow] (inputMath) -- (gru);
 \draw[arrow] (gru) -- node[right, font=\scriptsize] {$\mathbf{E}$} (attn);
 \draw[arrow] (attn) -- node[right, yshift=-0.2cm, font=\scriptsize]
  { $\tilde{\mathbf{E}} \in \mathbb{R}^{N_P \times d}$} (timeProj); \draw[arrow] (timeProj) -- node[right, font=\scriptsize]
  {$\mathbb{R}^{d \times N_L}$} (trans);
 \draw[arrow] (trans) -- (chanProj);
 \draw[arrow] (chanProj) -- (outputMath);

 \node [right=0.5cm of timeProj, font=\scriptsize, text=gray] {$N_P \to N_L$};
 \node [right=0.5cm of chanProj, font=\scriptsize, text=gray] {$d \to C$};

 \end{tikzpicture}

\caption{Overview of the proposed architecture.}
 \label{fig:dslh_arch}
\end{figure}
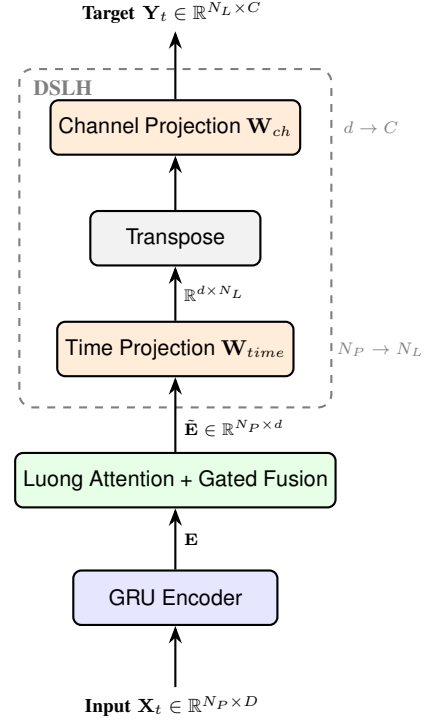

\subsection{GRU Encoder}

The core of our predictor is a multi-layer GRU network. Compared to LSTMs, GRUs
employ fewer gates, reducing parameter count while retaining the ability to model
temporal channel coherence. The GRU processes this sequence element-wise, where
$\vect{x}_\tau = \mat{X}_t[\tau,:] \in \mathbb{R}^{D}$ denotes the input feature
vector at the $\tau$-th position within the past window. The GRU maintains a hidden
state $\vect{h}_\tau \in \mathbb{R}^{d}$, where $d$ is the GRU hidden dimension.

For each time step $\tau \in \{1,\dots,\NP\}$, the GRU updates are
\begin{align}
 \vect{z}_\tau &= \sigma(\mat{W}_z \vect{x}_\tau + \mat{U}_z \vect{h}_{\tau-1} + \vect{b}_z), \\
 \vect{r}_\tau &= \sigma(\mat{W}_r \vect{x}_\tau + \mat{U}_r \vect{h}_{\tau-1} + \vect{b}_r), \\
 \tilde{\vect{h}}_\tau &= \tanh(\mat{W}_h \vect{x}_\tau + \mat{U}_h (\vect{r}_\tau \odot \vect{h}_{\tau-1}) + \vect{b}_h), \\
 \vect{h}_\tau &= (1 - \vect{z}_\tau) \odot \vect{h}_{\tau-1} + \vect{z}_\tau \odot \tilde{\vect{h}}_\tau,
\end{align}
where $\vect{z}_\tau \in \mathbb{R}^{d}$ and $\vect{r}_\tau \in \mathbb{R}^{d}$ are the update and reset gates, respectively; $\tilde{\vect{h}}_\tau \in \mathbb{R}^{d}$ is the candidate hidden state; $\sigma(\cdot)$ is the element-wise sigmoid; $\tanh(\cdot)$ is the element-wise hyperbolic tangent; and $\odot$ denotes the Hadamard (element-wise) product.
The trainable parameters are $\mat{W}_z,\mat{W}_r,\mat{W}_h \in \mathbb{R}^{d\times D}$, $\mat{U}_z,\mat{U}_r,\mat{U}_h \in \mathbb{R}^{d\times d}$, and biases $\vect{b}_z,\vect{b}_r,\vect{b}_h \in \mathbb{R}^{d}$.

Stacking the hidden states yields the encoder output sequence
\begin{equation}
\mat{E} = [\vect{h}_1, \dots, \vect{h}_{\NP}]^\top \in \mathbb{R}^{\NP \times d},
\end{equation}

\subsection{Scaled Luong Attention with Gated Fusion}
To capture long-range dependencies without the quadratic cost of self-attention, we employ a global attention mechanism based on \emph{Luong dot-product attention}~\cite{Luong2015}. Unlike standard approaches that pool the sequence into a single vector, we compute an \emph{attention-refined sequence} $\tilde{\mat{E}}$ that preserves the temporal structure required by the DSLH.

We use the final hidden state $\vect{h}_{\NP}$ as the query vector $\vect{q}$,
following the global attention formulation of Luong et al.~\cite{Luong2015}.
Scaled dot-product attention scores are computed as
\begin{equation}
\alpha_\tau = \frac{\vect{h}_\tau^\top \vect{q}}{\sqrt{d}}, \quad
w_\tau = \frac{\exp(\alpha_\tau)}{\sum_{j=1}^{\NP} \exp(\alpha_j)},
\end{equation}
where $\alpha_\tau$ denotes the unnormalized attention score measuring the
relevance of the $\tau$-th encoder hidden state to the query, and the scaling
factor $\sqrt{d}$ is adopted for numerical stability, as in Transformer-style
attention~\cite{Vaswani2017}. The resulting context vector
\begin{equation}
\vect{c} = \sum_{j=1}^{\NP} w_j \vect{h}_j
\end{equation}
summarizes the relevant history under the Luong global attention scheme.

A \emph{Bottleneck Gated Fusion} module~\cite{SENet2018,GLU2017} integrates 
this context back into the sequence with minimal parameter overhead:
\begin{align}
\vect{u}_\tau &= [\vect{h}_\tau; \vect{c}], \\
\vect{g}_\tau &= \sigma\!\left(\mat{W}_{2} \operatorname{ReLU}(\mat{W}_{1} \vect{u}_\tau)\right), \\
\vect{h}'_\tau &= \vect{g}_\tau \odot \vect{h}_\tau
+ (1 - \vect{g}_\tau) \odot \vect{c},
\label{eq:gfusion}
\end{align}
where $\mat{W}_{1} \in \mathbb{R}^{\frac{d}{r} \times 2d}$ and
$\mat{W}_{2} \in \mathbb{R}^{d \times \frac{d}{r}}$ constitute the bottleneck
with reduction ratio $r$. Here,
$\operatorname{ReLU}(\cdot)$ is the Rectified Linear Unit; and
$\vect{g}_\tau \in \mathbb{R}^{d}$ is the learnable fusion gate that dynamically
balances local temporal features and global context.
Because \eqref{eq:gfusion} produces $\vect{h}'_\tau$ for every encoder step
$\tau$, the model passes a refined sequence
$\tilde{\mat{E}} \in \mathbb{R}^{\NP \times d}$ to the prediction head rather
than a pooled vector. This is structurally compatible with the DSLH, whose
time projection mixes encoder positions and whose channel projection maps
feature dimensions. Accordingly, the DSLH serves here as a compact
sequence-to-sequence output head rather than a generic replacement for any
dense predictor.

\subsection{Dimension-wise Separable Linear Head (DSLH)}
A conventional dense output head that directly maps the encoder output
$\tilde{\mat{E}} \in \mathbb{R}^{\NP \times d}$ to the future sequence
$\hat{\mat{Y}}_t \in \mathbb{R}^{\NL \times \Cdim}$ is parameter-heavy,
particularly when the prediction horizon $\NL$ and output dimension $\Cdim$
are large. We adopt the \emph{Dimension-wise Separable Linear Head} (DSLH)
proposed in LinFormer~\cite{LinFormer2025}, which factorizes the output mapping
into decoupled projections along the temporal and channel dimensions.

DSLH consists of two linear mappings:
\begin{itemize}
\item \emph{Time projection:} $\mat{W}_{\mathrm{time}} \in \mathbb{R}^{\NP \times \NL}$,
which linearly mixes the $\NP$ encoder positions to produce $\NL$ future
positions.
\item \emph{Channel projection:} $\mat{W}_{\mathrm{ch}} \in \mathbb{R}^{d \times \Cdim}$,
which maps the feature dimension $d$ to the output dimension $\Cdim$.
\end{itemize}

The prediction is computed in two steps. First, a temporal projection is
applied:
\begin{equation}
\mat{G} \triangleq \mat{W}_{\mathrm{time}}^{\top} \tilde{\mat{E}}
\in \mathbb{R}^{\NL \times d},
\end{equation}
which aggregates information from the past window into $\NL$ future-aligned
representations. Second, a channel projection is applied to each future step:
\begin{equation}
\hat{\mat{Y}}_t \triangleq \mat{G}\mat{W}_{\mathrm{ch}}
\in \mathbb{R}^{\NL \times \Cdim}.
\end{equation}
Equivalently, the overall mapping can be written as
\begin{equation}
\hat{\mat{Y}}_t
= \left(\tilde{\mat{E}}^\top \mat{W}_{\mathrm{time}}\right)^\top
\mat{W}_{\mathrm{ch}}.
\end{equation}

Compared to a dense mapping from $\mathbb{R}^{\NP d}$ to $\mathbb{R}^{\NL \Cdim}$
with $\NP d \NL \Cdim$ parameters, DSLH requires only $\NP \NL + d \Cdim$
parameters (excluding biases).

\section{Performance Evaluation}\label{sec:results}

We evaluate the proposed model under 3GPP TR~38.901 channel 
conditions against GRU, LSTM, and LinFormer baselines. All models use the same hidden dimension, the same DSLH head, and UE speed as an additional input to provide a controlled comparison under a common output interface.

\subsection{Data and Evaluation Protocol}\label{sec:dataset}

Channels are generated using QuaDRiGa v2.8.1-0 under standardized 3GPP TR~38.901 UMa/UMi LOS and NLOS scenarios~\cite{QuaDRiGa2019,3GPP38901}. 
Each run simulates a UE moving along a \SI{150}{m} linear trajectory at 30 snapshots/m for a \SI{3.5}{GHz} carrier. 
QuaDRiGa produces a channel tensor in $\mathbb{C}^{R \times T \times P \times T_s}$, where $P$ denotes the number of multipath components and $T_s$ the number of time snapshots. 

All experiments use a $2\times2$ MIMO setup ($R=T=2$), yielding $D=9$ input features per time step and $\Cdim=8$ real-valued output dimensions. 
The evaluation protocol is based on standardized 3GPP TR~38.901 urban macro (UMa) and urban micro (UMi) environments, with held-out test scenarios covering general mixed-condition evaluation, pedestrian mobility, challenging UMi NLOS propagation, high-speed mobility, and favorable UMa LOS conditions, as summarized in Table~\ref{tab:scenarios}. 
Sliding windows with $\NP=128$ and $\NL=8$ produce approximately 900k training and 200k validation windows from the mixed training set.

\begin{table}[H]
\centering
\caption{Training and testing scenarios used in Section~\ref{sec:results}.}
\label{tab:scenarios}
\footnotesize
\setlength{\tabcolsep}{5pt}
\renewcommand{\arraystretch}{1.1}
\begin{tabular}{lccc}
\toprule
\textbf{ID} & \textbf{Phase} & \textbf{Environment} & \textbf{Speed (km/h)} \\
\midrule
--   & Training & UMa/UMi (LOS/NLOS) & 3--60 \\
Scenario I  & Testing  & UMa LOS + UMi NLOS & 10--60 \\
Scenario II & Testing  & UMi (LOS/NLOS)     & 3--5 \\
Scenario III& Testing  & UMi NLOS           & 10--60 \\
Scenario IV  & Testing  & UMa (LOS/NLOS)     & 80--120 \\
Scenario V  & Testing  & UMa LOS            & 10--60 \\
\bottomrule
\end{tabular}
\end{table}

\subsection{Metrics and Experimental Setup}\label{sec:setup}

Prediction accuracy is evaluated using sample-weighted NMSE (dB) as the primary
metric, with per-frame MSE tracked at each future step $k=1,\ldots,\NL$.
Training minimizes a weighted MSE loss with decaying weights $w_k = k^{-1/2}$.
A 200-sample gap between training and validation segments prevents data leakage,
and normalization statistics are derived from training data only. All results
are averaged over 10 independent runs; tables report Mean~$\pm$~Std.

The key simulation parameters, model configurations, and training hyperparameters used across all experiments are summarized in Table~\ref{tab:params}.

\begin{table}[h]
\centering
\caption{Simulation and Hyperparameter Configuration}
\label{tab:params}
\begingroup
\footnotesize
\setlength{\tabcolsep}{12pt}
\begin{tabular}{@{}lrlr@{}}
\toprule
\textbf{Parameter} & \textbf{Value} & \textbf{Parameter} & \textbf{Value} \\
\midrule
Past Window ($N_P$)        & 128              & Hidden Dim ($d$)        & 256 \\
Future Horizon ($N_L$)     & 8                & Network Layers          & 3 \\
Train/Val Split            & 80/20\%          & FFN Multiplier$^{\dagger}$ & $4\times$ \\
Batch Size                 & 256              & Dropout Rate            & 0.3 \\
Max Epochs                 & 100              & Reduction Ratio         & 4 \\
Learning Rate              & $3\times10^{-4}$ & Patience                & 15 \\
Weight Decay               & $1\times10^{-4}$ & Random Seeds            & 10 \\
\bottomrule
\multicolumn{4}{@{}l@{}}

{\footnotesize $^{\dagger}$ Specific to LinFormer architecture.}
\end{tabular}
\endgroup
\end{table}

\subsection{Comparative Performance Evaluation with Benchmarks}\label{sec:complexity}
Table~\ref{tab:complexity} reports efficiency on an NVIDIA RTX A2000 
(batch size 256). At $N_P=128$ with matched dimensions, GRU baseline achieves up to $2.8\times$ higher throughput than LinFormer, 
while the proposed model attains competitive or better NMSE across 
several scenarios (Table~\ref{tab:all_results}).
\begin{table}[h]
\centering
\caption{Model Complexity, Efficiency, and Throughput }
\label{tab:complexity}
\footnotesize
\renewcommand{\arraystretch}{1.2}
\setlength{\tabcolsep}{3pt}
\begin{tabular}{lcccc}
\toprule
\textbf{Model} & \textbf{Params} & \textbf{Complexity} & \textbf{Throughput} & \textbf{Avg NMSE} \\
& \textbf{(M)} & (GFLOPs) & (samples/s) & (dB) \\
\midrule
LinFormer & 1.68 & 0.40 & 5,811.07 & -13.86 $\pm$ 0.26 \\
LSTM      & 1.59 & 0.41 & 12,738.40 & -11.73 $\pm$ 0.03 \\
GRU      &1.19 & 0.31 &16,143.51 & -12.33 $\pm$ 0.07 \\
Proposed & 1.24 & 0.32 & 13,162.90 & -13.84 $\pm$ 0.15 \\
\bottomrule
\end{tabular}
\end{table}

\begin{table*}[t]
\centering
\caption{Performance Benchmark (NMSE in dB) with Mean $\pm$ Std across Test Scenarios}
\label{tab:all_results}
\footnotesize
\setlength{\tabcolsep}{3pt}
\renewcommand{\arraystretch}{1.2}
\begin{tabular}{lccccc}
\toprule
\textbf{Model}& \textbf{Scenario I} & \textbf{Scenario II} & \textbf{Scenario III} & \textbf{Scenario IV} & \textbf{Scenario V} \\
\midrule
LSTM      & -12.87 $\pm$ 0.14 & -13.34 $\pm$ 0.22 & -7.86 $\pm$ 0.09 & -8.80 $\pm$ 0.36 & -15.77 $\pm$ 0.23 \\
GRU       & -13.43 $\pm$ 0.15 & -14.14 $\pm$ 0.15 & -8.52 $\pm$ 0.05 & -9.21 $\pm$ 0.21 & -16.36 $\pm$ 0.19 \\
LinFormer & -14.48 $\pm$ 0.22 & -16.06 $\pm$ 0.33 & \textbf{-10.07 $\pm$ 0.19} & \textbf{-10.95 $\pm$ 0.19} & -17.73 $\pm$ 0.60 \\
\textbf{Proposed} & \textbf{-14.60 $\pm$ 0.20} & \textbf{-16.26 $\pm$ 0.23} & -9.26 $\pm$ 0.16 & -10.81 $\pm$ 0.11 & \textbf{-18.27 $\pm$ 0.25} \\
\bottomrule
\end{tabular}

\end{table*}

Table~\ref{tab:all_results} shows that the proposed model attains the best mean NMSE in Scenario I, II, and V, indicating that its advantages are most pronounced in general, pedestrian, and LOS-dominant settings. In Scenario V, the proposed model outperforms LinFormer ($-18.27$ vs.\ $-17.73$~dB) while exhibiting lower variance ($\pm 0.25$ vs.\ $\pm 0.60$). In Scenario II, it also achieves the best performance with $-16.26$~dB compared to LinFormer’s $-16.06$~dB. By contrast, LinFormer performs best in Scenario III and narrowly in Scenario IV. Notably, Scenario IV lies outside the training speed range (80--120~km/h vs.\ 3--60~km/h), yet the proposed model remains competitive, trailing by only 0.14~dB ($-10.81$ vs.\ $-10.95$~dB).
\begin{figure}[h]
\centering
\includegraphics[width=\linewidth]{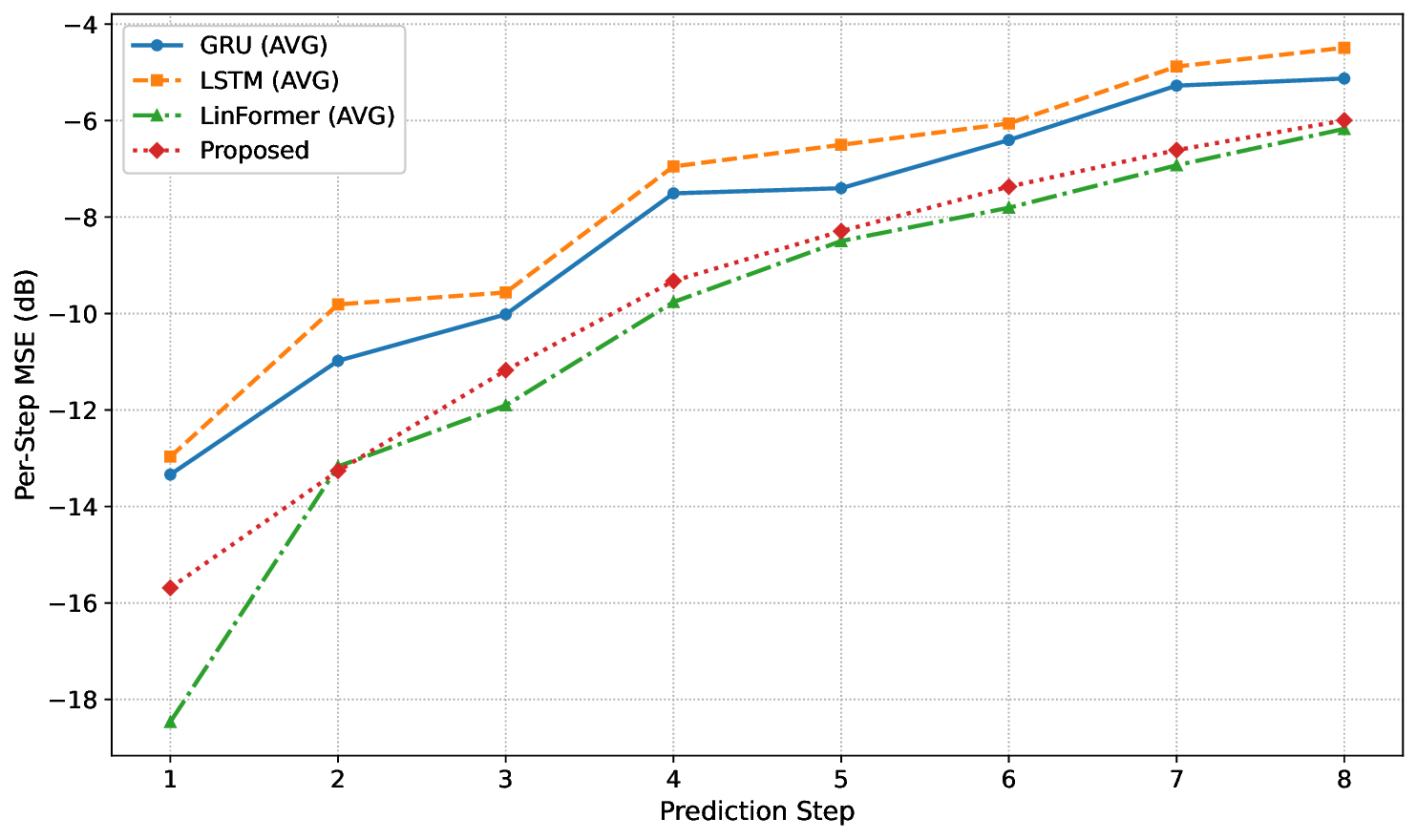}
\caption{NMSE over time (average across all test scenarios).}
\label{fig:nmse_horizon}
\end{figure}

Fig.~\ref{fig:nmse_horizon} illustrates the prediction error growth over the future horizon, averaged across all test scenarios. 
While LinFormer achieves the lowest error at the earliest prediction steps (Steps~1--3), the performance gap narrows significantly toward the later steps (Steps~4--8). 
Throughout the entire horizon, the proposed model remains consistently better than the GRU and LSTM baselines. 
Interpreted jointly with the throughput results in Table~\ref{tab:complexity}, the proposed model provides a highly favorable accuracy--latency trade-off, surpassing LinFormer in three out of five scenarios while running $2.3\times$ faster.
\subsection{Training Cost}

Table~\ref{tab:training_cost} compares training efficiency across models. Total training time reflects early stopping and therefore depends on both per-epoch cost and the number of epochs completed. The proposed model requires more training time than GRU (5431\,s vs.\ 3947\,s) due to the attention and gated fusion components, but substantially less than LinFormer (8126\,s) and at nearly half the GPU memory (567\,MB vs.\ 1093\,MB).
\begin{table}[H]
\centering
\caption{Training cost comparison across models}
\label{tab:training_cost}
\footnotesize
\setlength{\tabcolsep}{5pt}
\renewcommand{\arraystretch}{1.12}
\begin{tabular}{lccc}
\toprule
\textbf{Model} & \textbf{Train Time (s)} & \textbf{Average Epoch (s)} & \textbf{Mem (MB)} \\
\midrule
GRU     & $3947 \pm 1209$ & 54.22 & 543 \\
LSTM      & $2664 \pm 302$  & 59.45 & 520 \\
\textbf{Proposed}  & $5431 \pm 1338$ & 63.95 & 567 \\
LinFormer      & $8126 \pm 711$  & 83.84 & 1093 \\
\bottomrule
\end{tabular}
\begin{flushleft}
\footnotesize{All training experiments were conducted on an NVIDIA A40, while inference benchmarks were performed on an NVIDIA RTX A2000.}
\end{flushleft}
\end{table}
\subsection{Ablation Study}
Table~\ref{tab:ablation} supports two conclusions. First, replacing
fixed linear per-step fusion with gated fusion improves NMSE from
$-13.18$~dB to $-13.84$~dB while reducing parameters from $1.32$M to
$1.24$M, indicating that learned data-dependent fusion is more effective
than a fixed linear projection of the hidden and context vectors in this
architecture. Second, DSLH alone on top of a plain GRU is not sufficient,
as GRU (DSLH) underperforms GRU (Dense) ($-12.33$~dB vs.\ $-13.12$~dB).
However, once the sequence representation is strengthened by attention and
gated fusion, DSLH becomes effective, and the full proposed model
outperforms GRU (Dense) while yielding the best accuracy--parameter
trade-off.
\begin{table}[t]
\centering
\caption{Ablation study of architectural components}
\label{tab:ablation}
\footnotesize
\begin{threeparttable}
\setlength{\tabcolsep}{5pt}
\renewcommand{\arraystretch}{1.12}
\begin{tabular}{lcc}
\toprule
\textbf{Variant} & \textbf{Params (M)} & \textbf{Avg. NMSE (dB)} \\
\midrule
GRU (Dense)\tnote{a}               & 3.28 & $-13.12 \pm 0.15$ \\
GRU (DSLH)                         & 1.19 & $-12.33 \pm 0.07$ \\
GRU+Attn (Dense)\tnote{a}          & 3.42 & $-13.25 \pm 0.21$ \\
GRU+Attn+Gated Fusion (Dense)\tnote{a} & 3.33 & $-13.31 \pm 0.13$ \\
GRU+Attn (DSLH)\tnote{b}           & 1.32 & $-13.18 \pm 0.27$ \\
\textbf{Proposed}\tnote{c}         & \textbf{1.24} & $\mathbf{-13.84 \pm 0.15}$ \\
\bottomrule
\end{tabular}
\begin{tablenotes}[flushleft]
\footnotesize
\item[a] Dense denotes a direct mapping from the full refined sequence to the $N_L \times C$ output.
\item[b] GRU+Attn (DSLH) uses fixed linear per-step fusion $W[h_\tau; c]$.
\item[c] Proposed uses gated fusion.
\end{tablenotes}
\end{threeparttable}
\vspace{0.05cm}
\end{table}

Table~\ref{tab:ablation_speed} shows that appending UE speed consistently improves
NMSE across all scenarios, with gains ranging from 0.65~dB (Scenario IV) to 1.72~dB
(Scenario III). The largest improvements are observed in NLOS and LOS scenarios (Scenario III and V), indicating that speed provides useful information that enhances the model’s performance.

\begin{table}[t]
\centering
\caption{Impact of UE speed input on NMSE (dB) across Test scenarios}
\label{tab:ablation_speed}
\footnotesize
\setlength{\tabcolsep}{6pt}
\renewcommand{\arraystretch}{1.12}
\begin{tabular}{lcc}
\toprule
\textbf{Scenario} & \textbf{Proposed} & \textbf{No Speed} \\
\midrule
Scenario I   & \textbf{$-14.60 \pm 0.20$} & $-13.78 \pm 0.31$ \\
Scenario II  & \textbf{$-16.26 \pm 0.23$} & $-14.91 \pm 0.30$ \\
Scenario III & \textbf{$-9.26 \pm 0.16$}  & $-7.54 \pm 0.20$ \\
Scenario IV   & \textbf{$-10.81 \pm 0.11$} & $-10.16 \pm 0.23$ \\
Scenario V   & \textbf{$-18.27 \pm 0.25$} & $-16.88 \pm 0.35$ \\
\bottomrule
\end{tabular}
\end{table}

\section{Conclusion}
\label{sec:conclusion}

This letter presented a resource-efficient CSI predictor based on a GRU backbone, gated attention fusion, and a Dimension-wise Separable Linear Head (DSLH). Under 3GPP TR~38.901 channel simulations, the proposed model achieved an average NMSE of $-13.84$~dB while reducing parameters by 26\% and increasing inference throughput by approximately $2.3\times$ relative to a dimension-matched LinFormer baseline. The results indicate that the proposed architecture offers a competitive accuracy--efficiency trade-off for short-horizon CSI prediction, with particularly favorable performance in several LOS and mixed-condition scenarios. These findings suggest that, under the evaluated short-horizon 3GPP setting,
lightweight recurrent models with structured output heads remain competitive
alternatives to the considered LinFormer baseline. Future work will extend the study to wideband OFDM settings.

\bibliographystyle{IEEEtran}
\bibliography{bib}

\end{document}